\documentclass[modern]{aastex701}
\usepackage{graphicx}
\usepackage{subcaption}
\usepackage{CJKutf8}

\received{May 4, 2026}
\revised{May 6, 2026}
\accepted{May 7, 2026}

\submitjournal{RNAAS}

\shorttitle{Brown Dwarf Radius Inflation}
\shortauthors{Mehrotra et al.}

\graphicspath{{./}{figures/}}

\begin{document}

\title{Evidence of Radius Inflation Based on 50 Transiting Brown Dwarfs and Low-mass Stellar Companions}

%\correspondingauthor{Chih-Chun Hsu}
%\email{chsu@northwestern.edu}

\author[0009-0000-8229-971X]{Aarushi Mehrotra}
\affiliation{Walter Payton College Preparatory High School, 1034 N. Wells Ave, Chicago, IL, 60610, USA}
\email{aarushi.mehrotra0@gmail.com}

\author[0000-0002-5370-7494]{Chih-Chun Hsu}
\affiliation{Center for Interdisciplinary Exploration and Research in Astrophysics (CIERA), Northwestern University,
1800 Sherman Ave, Evanston, IL 60201, USA}
\email{chsu@northwestern.edu}

\author[0000-0003-0774-6502]{Jason J. Wang
\begin{CJK*}{UTF8}{gbsn}(王劲飞)\end{CJK*}}
\affil{Center for Interdisciplinary Exploration and Research in Astrophysics (CIERA), Northwestern University,
1800 Sherman Ave, Evanston, IL 60201, USA}
\affil{Department of Physics and Astronomy, Northwestern University, 2145 Sheridan Rd, Evanston, IL 60208, USA}
\email{jason.wang@northwestern.edu}

\author[0000-0002-9807-5435]{Christopher A. Theissen}
\affiliation{Department of Astronomy \& Astrophysics, University of California San Diego, La Jolla, CA 92093, USA}
\email{ctheissen@ucsd.edu}

\author[0000-0002-6523-9536]{Adam J.\ Burgasser}
\affiliation{Department of Astronomy \& Astrophysics, University of California San Diego, La Jolla, CA 92093, USA}
\email{aburgasser@ucsd.edu}

\begin{abstract}

Statistical assessment of stellar parameters enables validation and improvements in theoretical models. We compiled a sample of 85 transiting stellar and substellar companions, with masses ranging from $\sim$13--100 $M_\mathrm{Jup}$.
We focus on analyzing 50 transiting companions with robust ages and model constraints, and evaluate the degree
of radius inflation versus mass, separation, equilibrium temperature, and metallicity.
Our evaluation of the differences between measured and model radii indicates an $8.7\pm1.9$\% radius inflation for the full sample, at $4.6\sigma$ discrepancy, validating the existence of radius inflation at a population level in transiting brown dwarfs.
For brown dwarfs at separations $\leq 0.05$~au, we find an even higher radius inflation, with a median inflation of $16\pm6$\% at $2.7\sigma$, and the inflation decreases toward wider separations, likely due to reduced stellar irradiation. Finally, we provide our compilation for the community to use.

\end{abstract}


\section{Introduction} \label{sec:intro}

Brown dwarfs are substellar objects with masses less than 78.5 $M_{\text{J}}$ (0.075 M$_\mathrm{\odot}$; \citealp{Chabrier:2023aa}), bridging giant planets and lowest-mass stars.
Transiting brown dwarf companions provide empirical measurements of their masses and radii, which allow us to validate theoretical substellar evolutionary models.

Recent observations have identified radius inflation in low-mass stars and brown dwarfs, with measured radii being up to 25\% larger than model predictions in both transiting brown dwarfs \citep{Carmichael:2023aa}, 
isolated late-M dwarfs \citep{Hsu:2024aa}, and early-to-mid M dwarfs ($\sim$0.1--0.6 M$_\mathrm{\odot}$; \citealp{Kiman:2024aa}).
These studies used multiple methods, including transit method, projected radii, and SED fitting.
All three independent analyses found evidence of inflated radii, indicating unaccounted-for errors in very-low-mass stellar and substellar theoretical models.

For transiting brown dwarfs, individual studies reported inflated radii compared to theoretical models \citep{Carmichael:2023aa}.
However, a systematic and statistical analysis is required to assess the degree of radius inflation compared to the isolated brown dwarf population.
Fortunately, over the last few years, a flood of discoveries of transiting brown dwarf companions from TESS and ground-based facilities has made a robust statistical assessment now possible.
In this study, we assess the brown dwarf radii using the \textit{largest} literature compilation to date of transiting brown dwarf systems and compare them to the theoretical evolutionary models to evaluate the degree of radius inflation versus mass, separation, equilibrium temperature, and metallicity.
We also provide our compilation available on Zenodo derived from this study, for the community to use.

\section{Sample and Methods} \label{sec:observedsample}

We compiled a sample of 85 transiting brown dwarfs and low-mass stars from the literature.
Our compilations are drawn from several compilations in the literature, in particular, including 65 objects compiled from \cite{Barkaoui:2025aa}.
The full references are \cite{Pont:2005aa, Pont:2006aa, Deleuil:2008aa, Irwin:2010aa, Bouchy:2011aa, Siverd:2012aa, Triaud:2013aa, Diaz:2013aa, Tal-Or:2013aa, Moutou:2013aa, Diaz:2014aa, Bonomo:2015aa, Csizmadia:2015aa, Chaturvedi:2016aa, von-Boetticher:2017aa, Bayliss:2017aa, Nowak:2017aa, Shporer:2017aa, Gillen:2017aa, Irwin:2018aa, Canas:2018aa, Hodzic:2018aa, von-Boetticher:2019aa, Persson:2019aa, Zhou:2019ab, Carmichael:2019aa, David:2019aa, Jackman:2019aa, Parviainen:2020aa, Subjak:2020aa, Carmichael:2020aa, Mireles:2020aa, Grieves:2021aa, Carmichael:2021aa, Artigau:2021aa, Acton:2021aa, Benni:2021aa, Psaridi:2022aa, Gill:2022aa, Carmichael:2022aa, Sebastian:2022aa, Khandelwal:2023aa, Maldonado:2023aa, Schmidt:2023aa, Vowell:2023aa, Lin:2023aa, El-Badry:2023aa, Page:2024aa, Ferreira-dos-Santos:2024aa, Frost:2024aa, Henderson:2024aa, Henderson:2024ab, French:2024aa, Casewell:2024aa, Barkaoui:2025aa, Larsen:2025aa, Vowell:2025aa, Parsons:2025aa, Zhang:2026aa, Adams-Redai:2025aa}.

To ensure compatibility with the \cite{Baraffe:2003aa} model, we restricted our sample to objects with masses below $100$ M$_\mathrm{Jup}$ and ages less than 10~Gyr, which reduced our sample to 74 objects. We further constrained the sample to only have robust age uncertainties to ensure that radius differences were caused by evolutionary model limitations, further reducing our sample to 50 objects.
The masses of host stars range from 0.37--2.38 M$_\mathrm{\odot}$ and the temperatures range from $\sim$3350--15900 K.
The metallicities ([Fe/H]) of host stars range from $-$0.3 to $+$0.4~dex. 
Our literature sample have companion masses ranging from $\sim$13--100 M$_\mathrm{Jup}$ and radii ranging from $\sim$0.7--3.1 R$_\mathrm{Jup}$.
The separations in the sample range from 0.004--0.317~au, with 70\% of the sample having separations less than 0.1~au due to the sensitivity of the transit technique.
The majority of our sources are at field ages, with 37 older than 1~Gyr and 13 younger than 1~Gyr. 
We also collected equilibrium temperatures for 40 sources.

\section{Results and Discussion} \label{sec:results}

We compute the radius differences of sources in our sample by comparing the measured radii and the \cite{Baraffe:2003aa} model radii.
For each source, we sampled the observed radii under a Gaussian distribution.
We then inferred the model radii using observed masses (Gaussian distribution) and ages from host stars (uniform distribution).
The errors were propagated via 1000 Monte Carlo samplers.
We define the radius difference (in percentages) as $\frac{R_{\mathrm{Empirical}} - R_{\mathrm{Model}}}{R_{\mathrm{Model}}}$, and report in median and $\pm 1~\sigma$ uncertainties with Jackknife sampling.

\begin{figure}[ht!]
    \centering

    % first row
    \begin{subfigure}[b]{0.49\linewidth}
        \centering
        \includegraphics[width=\textwidth]{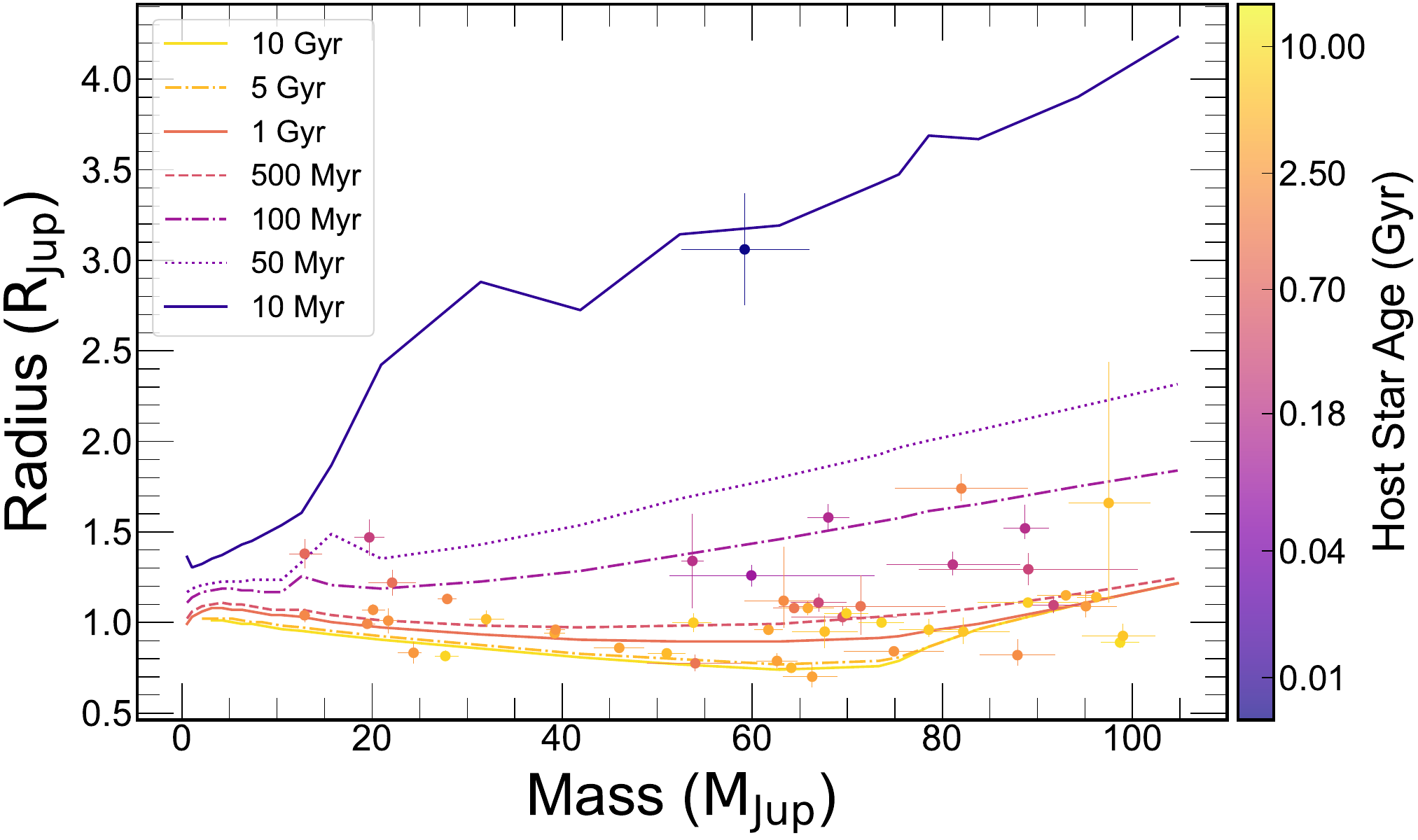}
        \caption{}
        \label{fig:fig1}
    \end{subfigure}
    \hfill
        \begin{subfigure}[b]{0.49\linewidth}
        \centering
        \includegraphics[width=\textwidth]{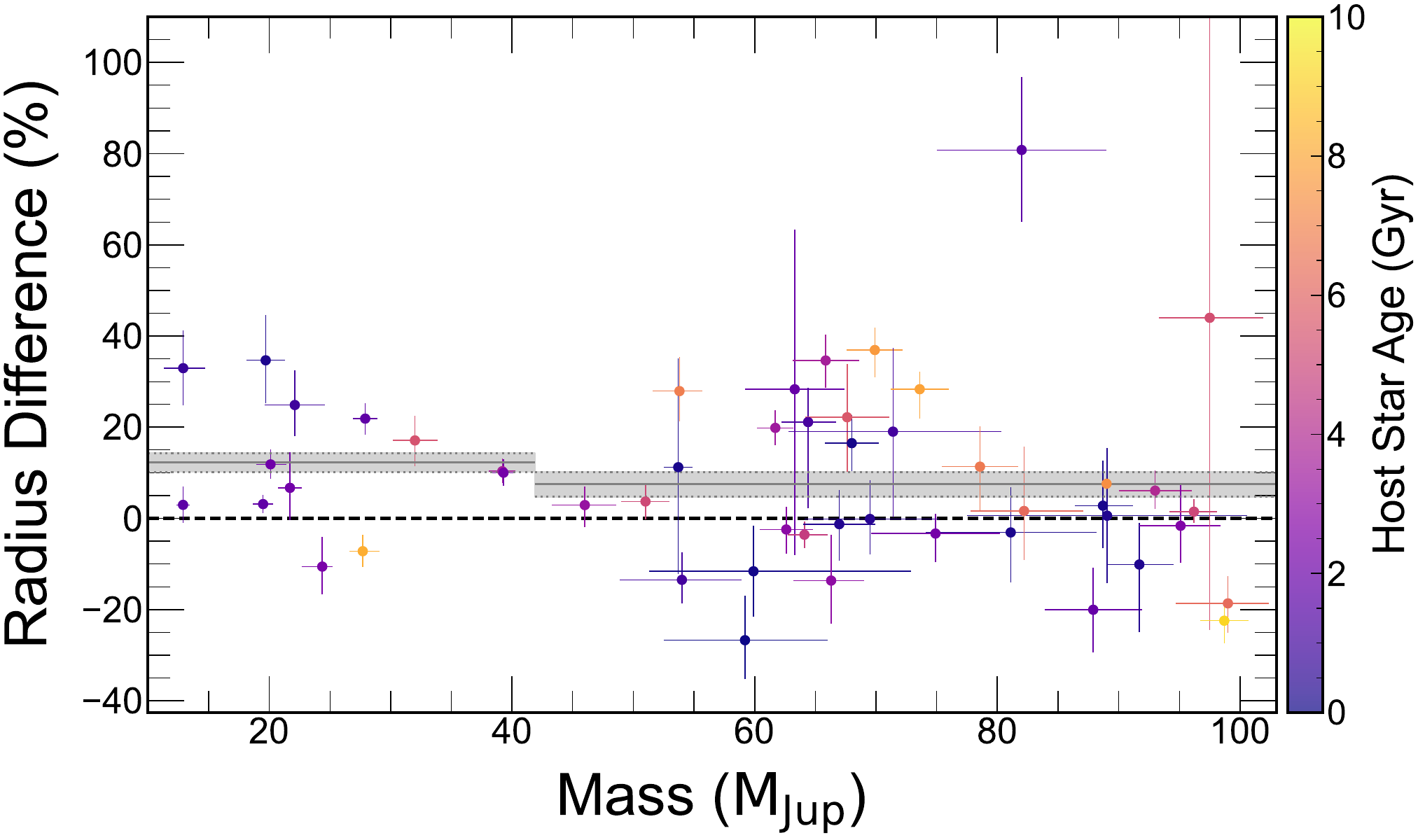}
        \caption{}
        \label{fig:fig2}
    \end{subfigure}

    \vskip\baselineskip  % vertical space between rows

    \vspace{-3mm}

    % second row
    \begin{subfigure}[b]{0.49\linewidth}
        \centering
        \includegraphics[width=\textwidth]{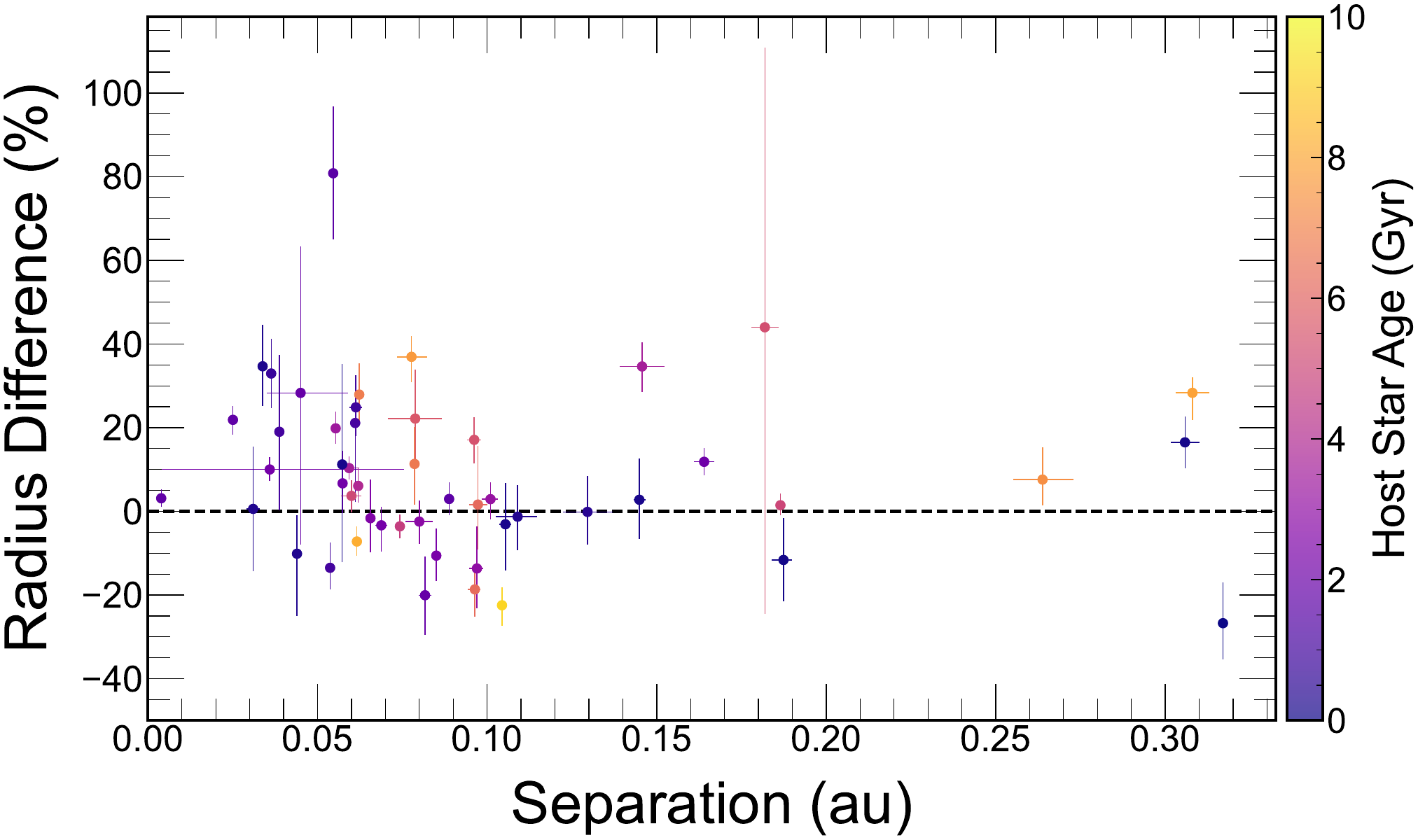}
        \caption{}
        \label{fig:fig3}
    \end{subfigure}
    \hfill
    \begin{subfigure}[b]{0.49\linewidth}
        \centering
        \includegraphics[width=\textwidth]{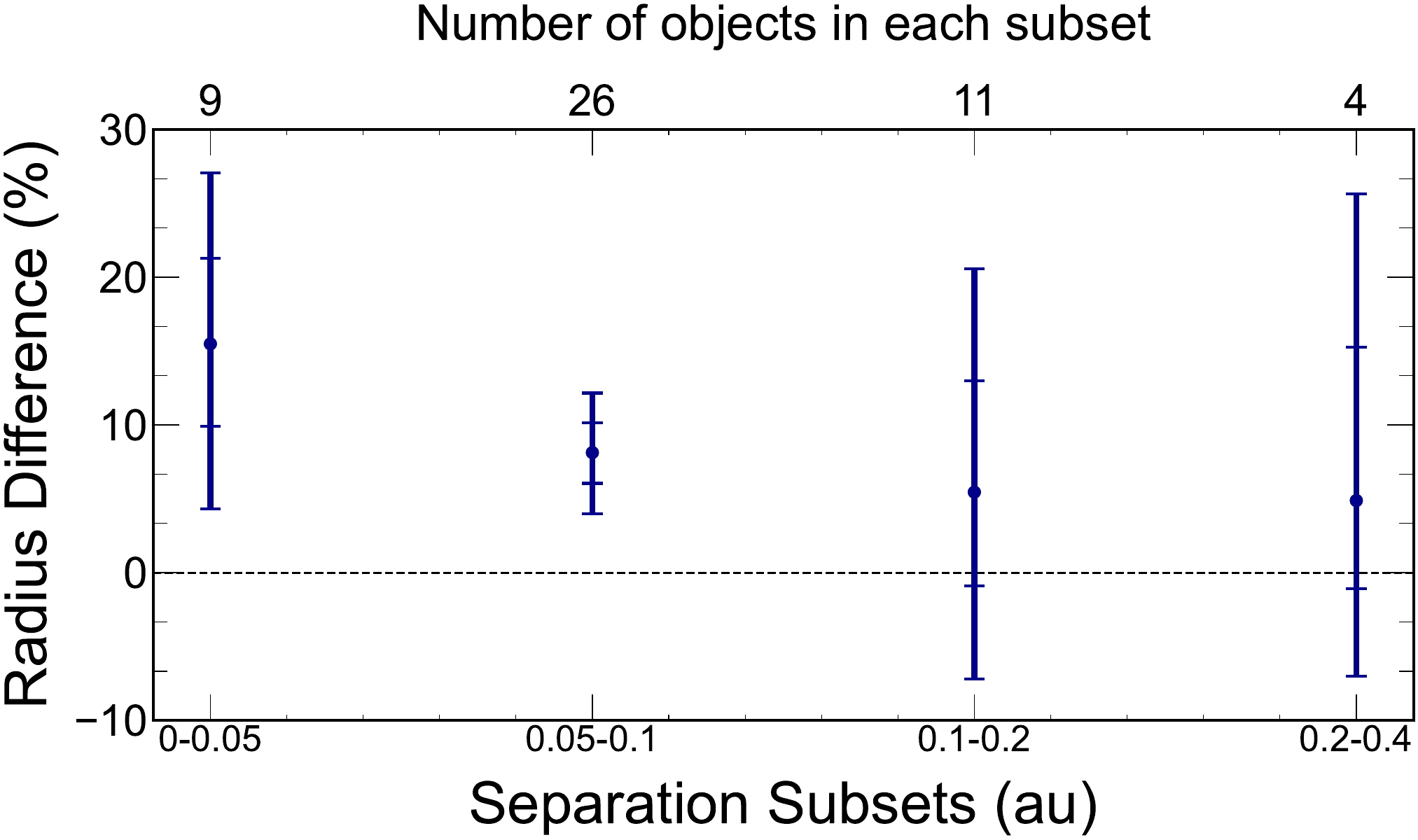}
        \caption{}
        \label{fig:fig4}
    \end{subfigure}

    \vspace{-3mm}

    \vskip\baselineskip  % vertical space between rows

    % third row
    \begin{subfigure}[b]{0.49\linewidth}
        \centering
        \includegraphics[width=1\linewidth]{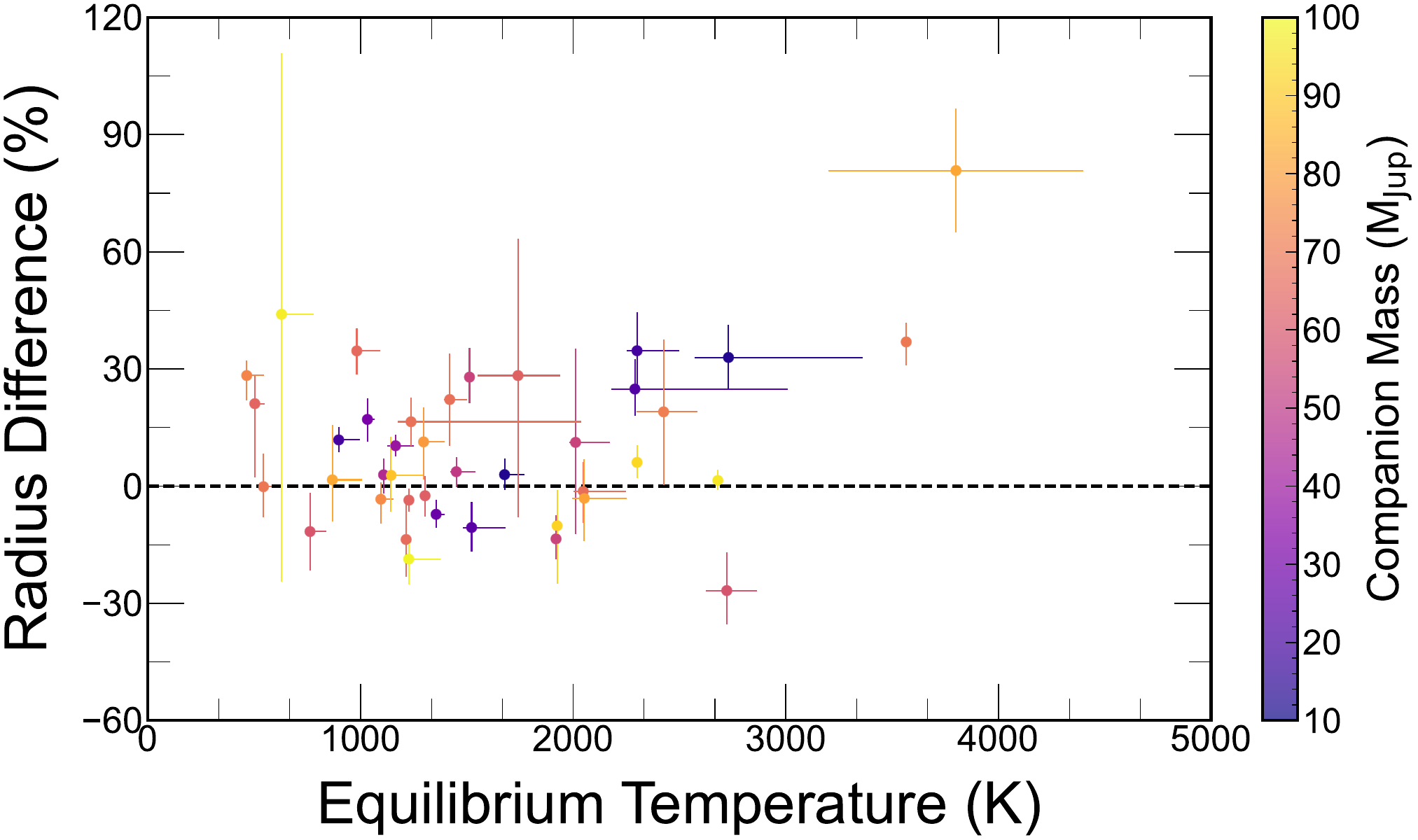}
        \caption{}
        \label{fig:fig5}
    \end{subfigure}
    \hfill
    \begin{subfigure}[b]{0.49\linewidth}
        \centering
        \includegraphics[width=\textwidth]{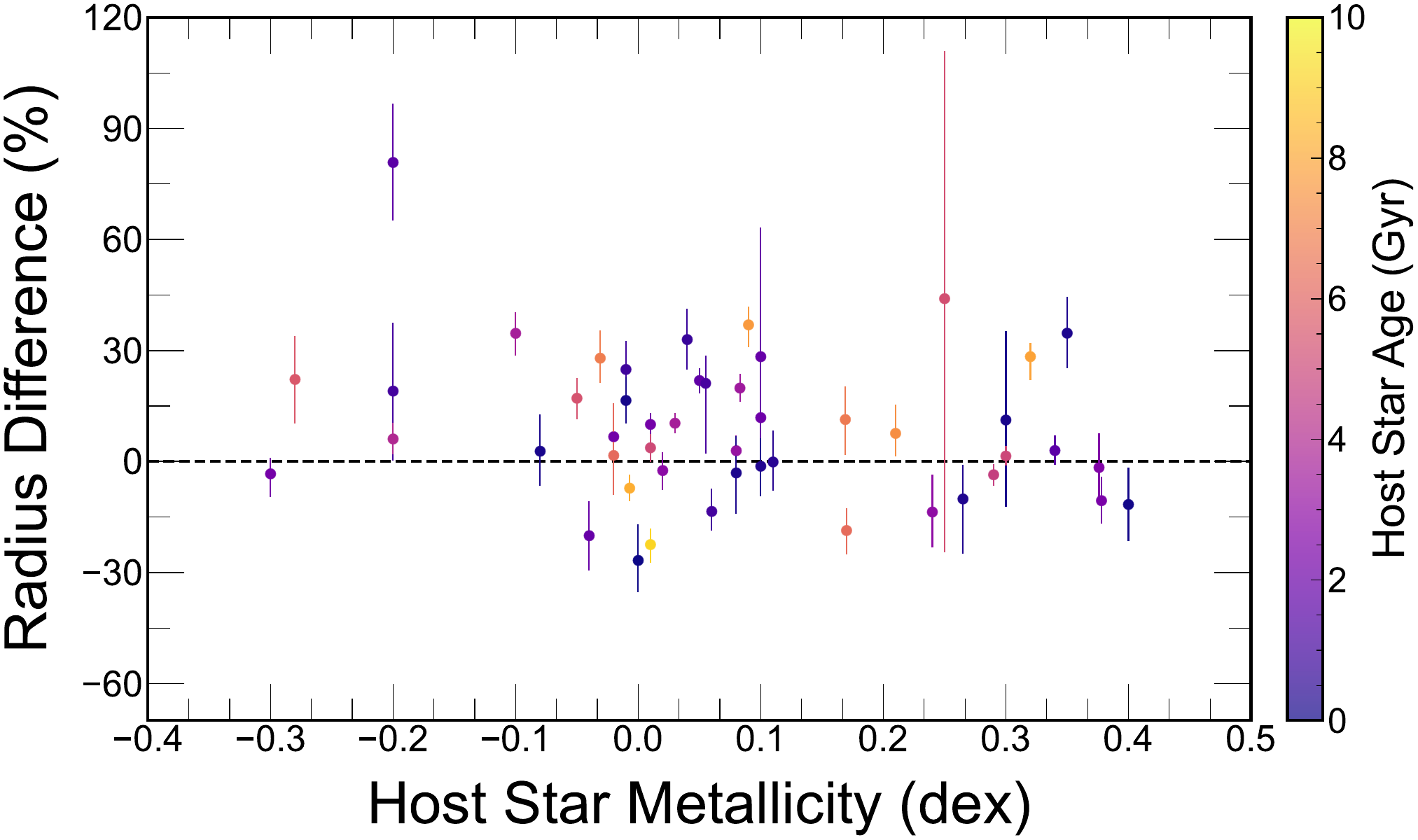}
        \caption{}
        \label{fig:fig6}
    \end{subfigure}

    \vspace{-0mm}
     
    \caption{
    (a) Radius versus mass, along with \cite{Baraffe:2003aa} evolutionary tracks, color-coded by ages;
    (b) Radius difference versus mass, along with median and 1~$\sigma$ radius differences by mass bins, color-coded by ages;
    (c) Radius difference versus orbital separation, color-coded by ages;
    (d) Median radius difference versus separation bins.
    The central point on each blue line represents the median difference, and
    each tick denotes 1$\sigma$ and 2$\sigma$ errors;
    (e) Radius difference versus equilibrium temperature, color-coded by their masses; 
    (f) Radius difference versus host star metallicity, color-coded by ages.
    Our compilation is available in the data behind this figure in the online journal.}
    \label{fig:1}
\end{figure}

Our overall distribution of radii and masses is shown in Figure~\ref{fig:fig1}.
%The measured radii are qualitatively consistent with the model expectations across different ages.
We then quantitatively compared the overall inflation across different models, including \cite{Baraffe:2003aa} and \cite{Burrows:2001aa}.
We found the observations show inflation compared to both models at 4.6~$\sigma$ and 5.2~$\sigma$, respectively.
The \cite{Baraffe:2003aa} model has slightly lower radius differences ($8.7\pm1.9$\%) than the \cite{Burrows:2001aa} models ($10.4\pm2.0$\%).

% 1b
The radius difference of observations and models across different masses are shown in Figure \ref{fig:fig2}. The majority (34 out of 50) of the companions in our sample exhibit inflated radii, with 13 of which being inflated by $>$3~$\sigma$.
% 1b
We find that radius inflation is larger in masses $M$=0.01--0.04~M$_{\odot}$ ($\Delta R/R \approx 12.2\pm{1.8}$\%) compared to masses $M$=0.04--0.1~M$_{\odot}$) ($\Delta R/R \approx {7.9}\pm{2.5}$\%).

% 1c & 1d
Figure \ref{fig:fig3} shows the distributions of radius differences versus separation.
In Figure~\ref{fig:fig4}, we binned our sample across different separations, from 0--0.05~au, 0.05--0.1~au, 0.1--0.2~au, and 0.2--0.4~au, and found radius differences of $16\pm{6}$\% (2.7$\sigma$), $8.3\pm1.7$\% (5$\sigma$), $6^{+7}_{-6}$\% (1$\sigma$), and $5^{+12}_{-5}$\% (1$\sigma$), respectively.
From closer to further separations of the host stars, the radius inflation decreases, which could be caused by irradiation \citep{Mukherjee:2025ab}.
However, a larger sample is required to validate our findings, especially for wider orbits $>0.1$~au.
We found no correlation of radius differences versus equilibrium temperature (Figure \ref{fig:fig5}), or host star metallicities (Figure \ref{fig:fig6}).
Future discoveries and characterization of transiting brown dwarfs of larger size, younger age, and lower metallicity will improve our understanding of substellar models, which can be placed with radii of isolated young brown dwarfs \citep{Hsu:2024aa}.

\textbf{The catalog compiled in this work is available in the data behind Figure~\ref{fig:1} and the Zenodo DOI: \href{https://doi.org/10.5281/zenodo.20019306}{10.5281/zenodo.20019306}.}

\software{
SPLAT \citep{Burgasser:2017ac}
}

\newpage

\bibliography{bibliography}{}

@article{Chabrier:2023aa,
  author =        {{Chabrier}, Gilles and {Baraffe}, Isabelle and
                   {Phillips}, Mark and {Debras}, Florian},
  journal =       {\aap},
  month =         mar,
  pages =         {A119},
  title =         {{Impact of a new H/He equation of state on the
                   evolution of massive brown dwarfs. New determination
                   of the hydrogen burning limit}},
  volume =        {671},
  year =          {2023},
  doi =           {10.1051/0004-6361/202243832},
  eid =           {A119},
}

@article{Carmichael:2023aa,
  author =        {{Carmichael}, Theron W.},
  journal =       {\mnras},
  month =         mar,
  number =        {4},
  pages =         {5177-5190},
  title =         {{Improved radius determinations for the transiting
                   brown dwarf population in the era of Gaia and TESS}},
  volume =        {519},
  year =          {2023},
  doi =           {10.1093/mnras/stac3720},
}

@article{Hsu:2024aa,
  author =        {{Hsu}, Chih-Chun and {Burgasser}, Adam J. and
                   {Theissen}, Christopher A. and {Birky}, Jessica L. and
                   {Aganze}, Christian and {Gerasimov}, Roman and
                   {Schmidt}, Sarah J. and {Blake}, Cullen H. and
                   {Covey}, Kevin R. and {Moreno-Hilario}, Elizabeth and
                   {Gelino}, Christopher R. and {Serna}, Javier and
                   {Brownstein}, Joel R. and {Cunha}, Katia},
  journal =       {\apjs},
  month =         oct,
  number =        {2},
  pages =         {40},
  title =         {{The Brown Dwarf Kinematics Project (BDKP). VI.
                   Ultracool Dwarf Radial and Rotational Velocities from
                   SDSS/APOGEE High-resolution Spectroscopy}},
  volume =        {274},
  year =          {2024},
  doi =           {10.3847/1538-4365/ad6b27},
  eid =           {40},
}

@article{Kiman:2024aa,
  author =        {{Kiman}, Rocio and {Brandt}, Timothy D. and
                   {Faherty}, Jacqueline K. and {Popinchalk}, Mark},
  journal =       {\aj},
  month =         sep,
  number =        {3},
  pages =         {126},
  title =         {{Accurate and Model-independent Radius Determination
                   of Single FGK and M Dwarfs Using Gaia DR3 Data}},
  volume =        {168},
  year =          {2024},
  doi =           {10.3847/1538-3881/ad5cf3},
  eid =           {126},
}

@article{Barkaoui:2025aa,
  author =        {{Barkaoui}, K. and {Sebastian}, D. and
                   {Z{\'u}{\~n}iga-Fern{\'a}ndez}, S. and
                   {Triaud}, A.~H.~M.~J. and {Rackham}, B.~V. and
                   {Burgasser}, A.~J. and {Carmichael}, T.~W. and
                   {Gillon}, M. and {Theissen}, C. and {Softich}, E. and
                   {Rojas-Ayala}, B. and {Srdoc}, G. and {Soubkiou}, A. and
                   {Fukui}, A. and {Timmermans}, M. and {Stalport}, M. and
                   {Burdanov}, A. and {Ciardi}, D.~R. and
                   {Collins}, K.~A. and {Davis}, Y.~T. and {Davoudi}, F. and
                   {de Wit}, J. and {Demory}, B.~O. and {Deveny}, S. and
                   {Dransfield}, G. and {Ducrot}, E. and {Florian}, L. and
                   {Gan}, T. and {G{\'o}mez Maqueo Chew}, Y. and
                   {Hooton}, M.~J. and {Howell}, S.~B. and
                   {Jenkins}, J.~M. and {Littlefield}, C. and
                   {Mart{\'\i}n}, E.~L. and {Murgas}, F. and
                   {Niraula}, P. and {Palle}, E. and {Pedersen}, P.~P. and
                   {Pozuelos}, F.~J. and {Queloz}, D. and {Ricker}, G. and
                   {Schwarz}, R.~P. and {Seager}, S. and {Shporer}, A. and
                   {Scott}, M.~G. and {Stockdale}, C. and {Winn}, J.},
  journal =       {\aap},
  month =         apr,
  pages =         {A44},
  title =         {{TOI-6508 b: A massive transiting brown dwarf
                   orbiting a low-mass star}},
  volume =        {696},
  year =          {2025},
  doi =           {10.1051/0004-6361/202453508},
  eid =           {A44},
}

@article{Pont:2005aa,
  author =        {{Pont}, F. and {Melo}, C.~H.~F. and {Bouchy}, F. and
                   {Udry}, S. and {Queloz}, D. and {Mayor}, M. and
                   {Santos}, N.~C.},
  journal =       {\aap},
  month =         apr,
  number =        {2},
  pages =         {L21-L24},
  title =         {{A planet-sized transiting star around OGLE-TR-122.
                   Accurate mass and radius near the hydrogen-burning
                   limit}},
  volume =        {433},
  year =          {2005},
  doi =           {10.1051/0004-6361:200500025},
}

@article{Pont:2006aa,
  author =        {{Pont}, F. and {Moutou}, C. and {Bouchy}, F. and
                   {Behrend}, R. and {Mayor}, M. and {Udry}, S. and
                   {Queloz}, D. and {Santos}, N. and {Melo}, C.},
  journal =       {\aap},
  month =         mar,
  number =        {3},
  pages =         {1035-1039},
  title =         {{Radius and mass of a transiting M dwarf near the
                   hydrogen-burning limit. OGLE-TR-123}},
  volume =        {447},
  year =          {2006},
  doi =           {10.1051/0004-6361:20053692},
}

@article{Deleuil:2008aa,
  author =        {{Deleuil}, M. and {Deeg}, H.~J. and {Alonso}, R. and
                   {Bouchy}, F. and {Rouan}, D. and {Auvergne}, M. and
                   {Baglin}, A. and {Aigrain}, S. and {Almenara}, J.~M. and
                   {Barbieri}, M. and {Barge}, P. and {Bruntt}, H. and
                   {Bord{\'e}}, P. and {Collier Cameron}, A. and
                   {Csizmadia}, Sz. and {de La Reza}, R. and
                   {Dvorak}, R. and {Erikson}, A. and {Fridlund}, M. and
                   {Gandolfi}, D. and {Gillon}, M. and {Guenther}, E. and
                   {Guillot}, T. and {Hatzes}, A. and {H{\'e}brard}, G. and
                   {Jorda}, L. and {Lammer}, H. and {L{\'e}ger}, A. and
                   {Llebaria}, A. and {Loeillet}, B. and {Mayor}, M. and
                   {Mazeh}, T. and {Moutou}, C. and {Ollivier}, M. and
                   {P{\"a}tzold}, M. and {Pont}, F. and {Queloz}, D. and
                   {Rauer}, H. and {Schneider}, J. and {Shporer}, A. and
                   {Wuchterl}, G. and {Zucker}, S.},
  journal =       {\aap},
  month =         dec,
  number =        {3},
  pages =         {889-897},
  title =         {{Transiting exoplanets from the CoRoT space mission .
                   VI. CoRoT-Exo-3b: the first secure inhabitant of the
                   brown-dwarf desert}},
  volume =        {491},
  year =          {2008},
  doi =           {10.1051/0004-6361:200810625},
}

@article{Irwin:2010aa,
  author =        {{Irwin}, Jonathan and {Buchhave}, Lars and
                   {Berta}, Zachory K. and {Charbonneau}, David and
                   {Latham}, David W. and {Burke}, Christopher J. and
                   {Esquerdo}, Gilbert A. and {Everett}, Mark E. and
                   {Holman}, Matthew J. and {Nutzman}, Philip and
                   {Berlind}, Perry and {Calkins}, Michael L. and
                   {Falco}, Emilio E. and {Winn}, Joshua N. and
                   {Johnson}, John A. and {Gazak}, J. Zachary},
  journal =       {\apj},
  month =         aug,
  number =        {2},
  pages =         {1353-1366},
  title =         {{NLTT 41135: A Field M Dwarf + Brown Dwarf Eclipsing
                   Binary in a Triple System, Discovered by the MEarth
                   Observatory}},
  volume =        {718},
  year =          {2010},
  doi =           {10.1088/0004-637X/718/2/1353},
}

@article{Bouchy:2011aa,
  author =        {{Bouchy}, F. and {Deleuil}, M. and {Guillot}, T. and
                   {Aigrain}, S. and {Carone}, L. and {Cochran}, W.~D. and
                   {Almenara}, J.~M. and {Alonso}, R. and {Auvergne}, M. and
                   {Baglin}, A. and {Barge}, P. and {Bonomo}, A.~S. and
                   {Bord{\'e}}, P. and {Csizmadia}, Sz. and
                   {de Bondt}, K. and {Deeg}, H.~J. and
                   {D{\'\i}az}, R.~F. and {Dvorak}, R. and {Endl}, M. and
                   {Erikson}, A. and {Ferraz-Mello}, S. and
                   {Fridlund}, M. and {Gandolfi}, D. and
                   {Gazzano}, J.~C. and {Gibson}, N. and {Gillon}, M. and
                   {Guenther}, E. and {Hatzes}, A. and {Havel}, M. and
                   {H{\'e}brard}, G. and {Jorda}, L. and {L{\'e}ger}, A. and
                   {Lovis}, C. and {Llebaria}, A. and {Lammer}, H. and
                   {MacQueen}, P.~J. and {Mazeh}, T. and {Moutou}, C. and
                   {Ofir}, A. and {Ollivier}, M. and {Parviainen}, H. and
                   {P{\"a}tzold}, M. and {Queloz}, D. and {Rauer}, H. and
                   {Rouan}, D. and {Santerne}, A. and {Schneider}, J. and
                   {Tingley}, B. and {Wuchterl}, G.},
  journal =       {\aap},
  month =         jan,
  pages =         {A68},
  title =         {{Transiting exoplanets from the CoRoT space mission.
                   XV. CoRoT-15b: a brown-dwarf transiting companion}},
  volume =        {525},
  year =          {2011},
  doi =           {10.1051/0004-6361/201015276},
  eid =           {A68},
}

@article{Siverd:2012aa,
  author =        {{Siverd}, Robert J. and {Beatty}, Thomas G. and
                   {Pepper}, Joshua and {Eastman}, Jason D. and
                   {Collins}, Karen and {Bieryla}, Allyson and
                   {Latham}, David W. and {Buchhave}, Lars A. and
                   {Jensen}, Eric L.~N. and {Crepp}, Justin R. and
                   {Street}, Rachel and {Stassun}, Keivan G. and
                   {Gaudi}, B. Scott and {Berlind}, Perry and
                   {Calkins}, Michael L. and {DePoy}, D.~L. and
                   {Esquerdo}, Gilbert A. and {Fulton}, Benjamin J. and
                   {F{\H{u}}r{\'e}sz}, G{\'a}bor and {Geary}, John C. and
                   {Gould}, Andrew and {Hebb}, Leslie and
                   {Kielkopf}, John F. and {Marshall}, Jennifer L. and
                   {Pogge}, Richard and {Stanek}, K.~Z. and
                   {Stefanik}, Robert P. and {Szentgyorgyi}, Andrew H. and
                   {Trueblood}, Mark and {Trueblood}, Patricia and
                   {Stutz}, Amelia M. and {van Saders}, Jennifer L.},
  journal =       {\apj},
  month =         dec,
  number =        {2},
  pages =         {123},
  title =         {{KELT-1b: A Strongly Irradiated, Highly Inflated,
                   Short Period, 27 Jupiter-mass Companion Transiting a
                   Mid-F Star}},
  volume =        {761},
  year =          {2012},
  doi =           {10.1088/0004-637X/761/2/123},
  eid =           {123},
}

@article{Triaud:2013aa,
  author =        {{Triaud}, A.~H.~M.~J. and {Hebb}, L. and
                   {Anderson}, D.~R. and {Cargile}, P. and
                   {Collier Cameron}, A. and {Doyle}, A.~P. and
                   {Faedi}, F. and {Gillon}, M. and
                   {Gomez Maqueo Chew}, Y. and {Hellier}, C. and
                   {Jehin}, E. and {Maxted}, P. and {Naef}, D. and
                   {Pepe}, F. and {Pollacco}, D. and {Queloz}, D. and
                   {S{\'e}gransan}, D. and {Smalley}, B. and
                   {Stassun}, K. and {Udry}, S. and {West}, R.~G.},
  journal =       {\aap},
  month =         jan,
  pages =         {A18},
  title =         {{The EBLM project. I. Physical and orbital
                   parameters, including spin-orbit angles, of two
                   low-mass eclipsing binaries on opposite sides of the
                   brown dwarf limit}},
  volume =        {549},
  year =          {2013},
  doi =           {10.1051/0004-6361/201219643},
  eid =           {A18},
}

@article{Bonomo:2015aa,
  author =        {{Bonomo}, A.~S. and {Sozzetti}, A. and {Santerne}, A. and
                   {Deleuil}, M. and {Almenara}, J.-M. and {Bruno}, G. and
                   {D{\'\i}az}, R.~F. and {H{\'e}brard}, G. and
                   {Moutou}, C.},
  journal =       {\aap},
  month =         mar,
  pages =         {A85},
  title =         {{Improved parameters of seven Kepler giant companions
                   characterized with SOPHIE and HARPS-N}},
  volume =        {575},
  year =          {2015},
  doi =           {10.1051/0004-6361/201323042},
  eid =           {A85},
}
\bibliographystyle{aasjournal}

\end{document}